\documentclass
[prl,onecolumn,tightenlines,10pt,superscriptaddress]{revtex4}%
\usepackage{amssymb}
\usepackage{amsfonts}
\usepackage{amsmath}
\usepackage{graphicx}%
\providecommand{\U}[1]{\protect\rule{.1in}{.1in}}
\providecommand{\U}[1]{\protect\rule{.1in}{.1in}}
\providecommand{\U}[1]{\protect\rule{.1in}{.1in}}
\providecommand{\U}[1]{\protect\rule{.1in}{.1in}}

\begin{document}
\title[Hearing the Shape of the Ising Model with the Help of a
   Programmable
      Superconducting-Flux Annealer]{Hearing the Shape of the Ising Model  with a \\
   Programmable
      Superconducting-Flux Annealer}
\author{Walter Vinci\footnote{w.vinci(at)ucl.ac.uk}}
\affiliation{London Centre for Nanotechnology, University College London, WC1E 6BT London, UK}
\author{Klas Markstr\"{o}m}
\affiliation{Department of Mathematics and Mathematical Statistics, Ume\aa \ University,
S-901 87 Ume\aa , Sweden}
\author{Sergio Boixo}
\affiliation{Information Sciences Institute and Ming-Hsieh Department of Electrical
Engineering, University of Southern California, Los Angeles, CA 90089,
USA}
\affiliation{Google, Venice Beach, CA 90292, U.S.A}
\author{Aidan Roy}
\affiliation{D-Wave Systems Inc., 100-4401 Still Creek Drive, Burnaby, BC V5C 6G9, Canada}
\author{Federico M. Spedalieri}
\affiliation{Information Sciences Institute and Ming-Hsieh Department of Electrical
Engineering, University of Southern California, Los Angeles, CA 90089, USA}
\author{Paul A. Warburton}
\affiliation{London Centre for Nanotechnology, University College London, WC1E 6BT London, UK}
\author{Simone Severini}
\affiliation{Department of Computer Science, and Department of Physics \& Astronomy,
University College London, WC1E 6BT London, UK}

\begin{abstract}

  Two objects can be distinguished if they have different measurable
  properties. Thus, distinguishability depends on the Physics of the
  objects. In considering graphs, we revisit the Ising model as a
  framework to define physically meaningful spectral invariants. In
  this context, we introduce a family of refinements of the classical
  spectrum and consider the quantum partition function. We demonstrate
  that the energy spectrum of the quantum Ising Hamiltonian is a
  stronger invariant than the classical one without refinements. For
  the purpose of implementing the related physical systems, we perform
  experiments on a programmable annealer with superconducting flux
  technology. Departing from the paradigm of adiabatic computation, we
  take advantage of a noisy evolution of the device to generate
  statistics of low energy states. The graphs considered in the
  experiments have the same classical partition functions, but
  different quantum spectra. The data obtained from the annealer
 distinguish non-isomorphic graphs  via information contained in the
  classical refinements of the functions but not via the differences in the quantum spectra.

\end{abstract}
\maketitle

\section{Introduction} 

Kac's~\cite{Kak66} question \textquotedblleft Can one
hear the shape of a drum?\textquotedblright\ is part of the scientific pop
culture~\cite{Wiki}. The technical side of the question concerns our ability
to completely specify\ the geometry of a domain from the eigenvalues of its
Laplacian. The question has been reinterpreted in the study of Schr\"{o}dinger
operators on metric graphs by Gutkin and Smilansky~\cite{Gutkin01} and
restated in Algebraic Graphs Theory as \textquotedblleft Which graphs are
determined by their spectrum?\textquotedblright\ by van Dam and Haemers~\cite{vanDam03}. (Through this work, the \emph{spectrum} of a matrix $M$,
denoted by $S_{M}$, is the set of its eigenvalues.) While we commonly
employ different types of matrices to encode the structure of graphs, none has
yet been shown to efficiently provide a \emph{complete graph invariant}%
, \emph{i.e.}, a parameter that does not change under a permutation of the vertex labels. The spectrum of the adjacency matrix, for
example, is a common invariant and easily seen to satisfy the
\textquotedblleft if\textquotedblright\ part of this statement; however, it is
not a complete invariant, given the fact that co-spectral non-isomorphic
graphs are abundant~\cite{Godsil82a,Godsil82b} (see for instance Supplementary Information Section A). In the same spirit, physical scenarios
have suggested various notions of refined spectra as a tool for distinguishing
graphs, with partial degrees of success~\cite{Audenaert07a,Audenaert07b,Audenaert07b}. A
common intersection for these approaches is Quantum Mechanics, arguably due to
the popularization of quantum dynamics on graphs at the beginning of the last
decade~\cite{Konno08}.

It is interesting, not only from the historical point of view, to
observe that the strong link between Physics and graphs is via
the Ising model, perhaps the most studied model in Statistical Mechanics.
Originally proposed in 1925~\cite{Ising25} as a simplified description of the
magnetic properties of materials, the Ising model has found a vast number of
applications from Biology to Solid State Physics. Its great importance is
emphasized by exact solutions and numerical techniques for the identification
of phase transitions and critical phenomena~\cite{Huang90}. The Ising model
framework seems particularly suitable to observe differences between Classical
and Quantum Mechanics in terms of spectral information, since the quantum case
is directly obtained by adding an appropriate (transverse)\ magnetic field to
the classical Hamiltonian.

In what follows, we map a graph into an Ising model and interpret its
energy spectrum as a graph invariant, before and after the
\textquotedblleft switch\textquotedblright\ from Classical to Quantum
Mechanics. We demonstrate with exhaustive numerical examples that the
quantum spectrum is a stronger invariant and propose a general
framework to define physically meaningful graph
polynomials. Determining whether the quantum energy spectrum is a
complete invariant remains an open problem. We perform experiments on
a programmable annealer with superconducting flux technology~\cite{Johnson10}.  Our purpose is to \textquotedblleft hear the shape
of an Ising model\textquotedblright, by generating statistics of low
energy states as the outcome of a noisy evolution. The experiment is
run disregarding whether or not the state of the device follows an
adiabatic path along its instantaneous ground state, therefore against
the prescription for successful annealing~\cite{Santoro92}. In other
words, we are not only interested in the ground state but,
unconventionally, in the full output of a noisy computation. We obtain
data on non-isomorphic graphs that are distinguished by their quantum
energy spectra but not by the classical ones. 

\section{Results}
{\bf Classical Cospectrality and Ising models.} The Hamiltonian of the \emph{Ising
model}~\cite{Cipra87} (or, equivalent, $2$\emph{-state Potts model}) on a
graph $G$, with $n$ vertices $V(G)$ and edges $E(G)$, is defined by the
diagonal matrix 
\begin{equation}
H(G,J):=J\sum_{\{i,j\}\in E(G)}H(i,j) \equiv J\sum_{\{i,j\}} [A(G)]_{i,j} H(i,j)
\label{eq:IsingHamiltonian}
\end{equation}
where, for each edge
$\{i,j\}$, $H(i,j):=\bigotimes_{k=1}^{n}H(k)$ is a $2^{n}\times2^{n}$ matrix,
with $H(k)=\sigma_{z}$ if $k=i,j$ and $H(k)=I$, otherwise.  $A(G)$ is the \emph{adjacency matrix}
with $[A(G)]_{i,j}=1$ if $\{i,j\}\in E(G)$ and $[A(G)]_{i,j}=0$, otherwise.  $\sigma_{z}$
is the Pauli matrix in the $z$-th coordinate axis, $I$ is the identity matrix,
and $J$ is the strength of interaction.  From now on, whenever the interaction strength is not expressly indicated as, \emph{e.g.}, in $H(G)$, we implicitly
set $J=1$ for all edges. The \emph{partition function} of the Ising model on
$G$ is
\begin{equation}
Z(G,v)=\mathrm{Tr}(e^{-\beta H(G)}), \label{par}%
\end{equation}
where $\beta:=(k_{B}T)^{-1}$ and $v=e^{\beta J}-1$; $k_{B}$ is  Boltzmann's
constant, $T\in\mathbb{R}^{\geq0}$ is the temperature. By the
Fortuin-Kasteleyn~\cite{Fortuin72} combinatorial identity, $Z(G,v)$ is an
evaluation of the \emph{Tutte polynomial}~\cite{Tutte54a,Tutte54b}, %
which  is a fundamental invariant that determines many parameters
including girth, chromatic number, \emph{etc}. Remarkably, the Jones
polynomial of a knot is contained in the Tutte polynomial~\cite{Welsch93}.
Recall that, formally, $G$ and $G^{\prime}$ are \emph{isomorphic} if they are
the same graph up to a relabeling of the vertices. This is denoted by $G\cong
G^{\prime}$. It is not hard to find  graphs with the same Tutte polynomial (T-equivalent) that are  not isomorphic~\cite{Mier04a,Mier04b}:\ for example, all trees on the same number of vertices.

Observe that two graphs $G$ and $G^{\prime}$ have the same partition function if and only if they share the same spectrum of the Hamiltonian in Eq.~(\ref{eq:IsingHamiltonian}) (\emph{i.e.}  $Z(G,v)=Z(G^{\prime},v) \Leftrightarrow S_{H(G)}=S_{H(G^{\prime
})}$). We say that $G$ and $G^{\prime}$
are \emph{co-Ising} if $S_{H(G)}=S_{H(G^{\prime})}$. Since the Tutte polynomial is a generalization of the partition function, if two graphs
are T-equivalent then they share the same energy spectrum and thus are co-Ising. Thus, we know the following:%
\begin{eqnarray}
G\cong G^{\prime}\Rightarrow S_{H(G)}=S_{H(G^{\prime})} \nonumber \\
S_{H(G)}=S_{H(G^{\prime})}\nRightarrow G\cong G^{\prime}.
\end{eqnarray}

Intuitively, we may attempt a refinement by adding a longitudinal field. The
Hamiltonian of the \emph{Ising model on }$G$ \emph{with longitudinal field} is
defined by the diagonal matrix 
\begin{equation}
H_{L}(G,J,h):=H(G,J)+hM,
\end{equation} 
where $M:=\sum_{i=1}^{n}K(i)$ is a $2^{n}\times2^{n}$ matrix, with $K(i)=\bigotimes
_{k=1}^{n}H_{L}(k)$, $H_{L}(k)=\sigma_{z}$ if $k=i$ and $H_{L}(k)=I$ 
otherwise. Physically $h M$ can be interpreted as a constant external magnetic field applied to all
vertices. Again, we set $J=1$ and $h=1$ unless they are explicitly indicated.
We say that two graphs $G$ and $G^{\prime}$ are \emph{longitudinal field
co-Ising} if $S_{H_{L}(G,J,h)}=S_{H_{L}(G^{\prime},J,h)}$ for all values of
$J$ and $h$. The following equation summarizes what we know about graphs with
this property (see Supplementary Information Section A and B for examples):%
\begin{align}
\stackrel{\forall J,h} {S_{H_{L}(G,J,h)}   \; =\; S_{H_{L}(G^{\prime},J,h)}} &\Rightarrow \stackrel{\forall J} {S_{H(G,J)}%
=S_{H(G^{\prime},J)}\nonumber }\\
S_{H_{L}(G)}   =S_{H_{L}(G^{\prime})}&\nLeftrightarrow S_{H(G)}%
=S_{H(G^{\prime})}\label{exam} \nonumber \\
{\forall J,h} \;{S_{H_{L}(G,J,h)}    = S_{H_{L}(G^{\prime},J,h)}} &\nRightarrow G\cong G^{\prime}.
\end{align}

From the diagonal matrices $H(G)$ and $M$, we can define the \emph{energy} and
\emph{magnetization} vectors as $\mathbf{e}_{\sigma}(G)=H(G)_{\sigma,\sigma}$
and $\mathbf{m}_{\sigma}=M_{\sigma,\sigma}$, where $\sigma=0,2,...,2^{n}-1$
runs over the classical states of the Ising model on $G$, where $0$ denotes
the ground state. With the use of these vectors, the \emph{bivariate Ising
polynomial} is defined as~\cite{Andren09}:
\begin{equation}
Z(G;x,y)=\sum_{\sigma}x^{\mathbf{e}_{\sigma}(G)}y^{\mathbf{m}_{\sigma}}.
\label{bi}%
\end{equation}
Notice that the spectrum $S_{H_{L}(G,J,h)}$ can be obtained from
$Z(G;x,y)$ for all values of the constants $J$ and $h$, since a change
in these parameters is just a rescaling of the coefficients $x$ and
$y$. The Ising polynomial generalizes the partition function in
Eq.~\eqref{par} because $Z(G,e^{-J\beta},1)=Z(G,e^{J\beta}-1)\,$,
encodes the matching polynomial, is related to the van der Waerden
polynomial, and is contained in a more general polynomial introduced
by Goldberg, Jerrum and Paterson~\cite{Jerrum03a,Jerrum03b,Jerrum03c}. The bivariate Ising
polynomial in Eq.~\eqref{bi} can be intuitively generalized by working
with \emph{any} physical observable in addition to energy and
magnetization. If we denote by $\mathbf{o}_{\sigma}^{k}$ the
eigenvalues of a diagonal matrix (or observable) $\Lambda^{k}$, we can
then define a multivariate polynomial
\begin{equation}
Z(G;x,y,z_{k})=\sum_{\sigma}x^{\mathbf{e}_{\sigma}(G)}y^{\mathbf{m}_{\sigma}}%
{\displaystyle\prod\limits_{k}}
z_{k}^{\mathbf{o}_{\sigma}^{k}}. \label{mu}%
\end{equation}
An example is given by the (permutationally invariant) spin-glass order
parameter used by Hen and Young~\cite{Hen12}.
\\

{\bf Quantum Cospectrality.}  The invariants that we have so far considered belong to Classical
Physics.  We can now move into a quantum mechanical regime by adding a
further field. The Hamiltonian of the \emph{quantum Ising model} on
$G$, as proposed by Lieb, Schultz, and Mattis~\cite{Lieb61} (see also~\cite{Sachdev99}) is defined by the matrix%

\begin{equation}
H_{T}(G,J,h,\Delta):=H_{L}(G,J,h)+\Delta M_{T},
\end{equation}
where $\Delta\in\mathbb{R}$ is a transverse external magnetic field; here
$M_{T}:=\sum_{i=1}^{n}T(i)$ is a $2^{n}\times2^{n}$ matrix, with
$T(i)=\bigotimes_{k=1}^{n}H_{T}(k)$,  $H_{T}(k)=\sigma_{x}$ if $k=i$ and
$H_{T}(k)=I$ otherwise. As in the longitudinal case, $M_{T}$ does not depend
on $G$. Two graphs $G$ and $G^{\prime}$ are said to be \emph{quantum
co-Ising} if $S_{H_{T}(G,J,h,\Delta)}=S_{H_{T}(G^{\prime},J,h,\Delta)}$ for
all values of $J$, $h$ and $\Delta$. It follows from the definition that two graphs are quantum co-Ising if they are isomorphic. The quantum partition function is defined
analogously to the classical one:
\begin{equation}
Z_{T}(G,\beta,J,h,\Delta)=\mathrm{Tr}(e^{-\beta H_{T}(G,J,h,\Delta)}).
\end{equation}
Two graphs are quantum co-Ising if and only if they have the same quantum
partition function. The \textquotedblleft if\textquotedblright\ part of
this statement comes directly from the definition. For the \textquotedblleft
only if\textquotedblright\ part, observe that in the limit $\beta
\rightarrow\infty$, $Z_{T}(G)\simeq\nu_{0}e^{-\beta E_{0}}$
determines the lowest eigenvalue $E_{0}$ with its multiplicity $\nu_{0}$.
Similarly, in the same limit $Z_{T}(G)e^{\beta E_{0}}/\nu_0$ determines the value and multiplicity of
the second smallest eigenvalue. The whole spectrum is obtained iteratively.   The statement above and its proof are valid only for systems of finite size. It is a well-known fact that different Hamiltonians can have the same partition function in the thermodynamic limit.

We tested numerically the converse of this fact by computing the smallest
eigenvalue for $h=J=\Delta=1$. We tested all graphs with $n\leq9$, all bipartite graphs with $n\leq11$, all vertex
transitive graphs with $n\leq15$, all regular graphs with $n\leq11$, and all
trees with $n\leq14$ (also considered in~\cite{Andren09}). We failed to find a
counterexample. Hence,%
\begin{eqnarray}
G   \cong G^{\prime}\Rightarrow S_{H_{T}(G)}=S_{H_{T}(G^{\prime})}  \nonumber \\ 
S_{H_{T}(G)}   =S_{H_{T}(G^{\prime})}\overset{?}{\Rightarrow}G\cong
G^{\prime}.
\end{eqnarray}
The transverse field Ising Hamiltonian is a sum of non-commuting terms and
determining its full spectrum requires the diagonalization of a $2^{n}%
\times2^{n}$ matrix. We cannot generalize directly the quantum partition
function to a generating Ising polynomial as done in the classical case -- 
when eigenvalues are integers (for $J=h=1$) -- although we can
use the well-known Suzuki-Trotter formalism to obtain a classical
approximation~\cite{Dutta}; the direct calculation of the eigenvalues is
notoriously expensive, due to the size of the problem, and prone to errors,
making it difficult to numerically show the existence of non-isomorphic
quantum co-Ising graphs. A reasonable first approximation for this task is to
compute the absolute largest eigenvalue. That is what we have done in our
tests. Taking into account such difficulties, finding non-isomorphic quantum
co-Ising graphs is an open problem. Natural candidates are graphs for which
isomorphism testing is known to be harder to solve (\emph{e.g.}, graphs for
which the Weisfeiler-Lehman algorithm fails)~\cite{Arvind05}. Nevertheless, we
emphasize that spectral information provided by Quantum Mechanics is more
accurate than Classical Mechanics. It must be said that there are only a few
precise (and in fact negative)\ statements about the physically inspired graph
invariants which have been introduced recently (see~\cite{Audenaert07a,Audenaert07b,Audenaert07c, Hen12}
and the references therein) and that purely numerical analysis does not
guarantee sufficient generality.
\\

\begin{figure}
\centering
\includegraphics[width=0.95\columnwidth]{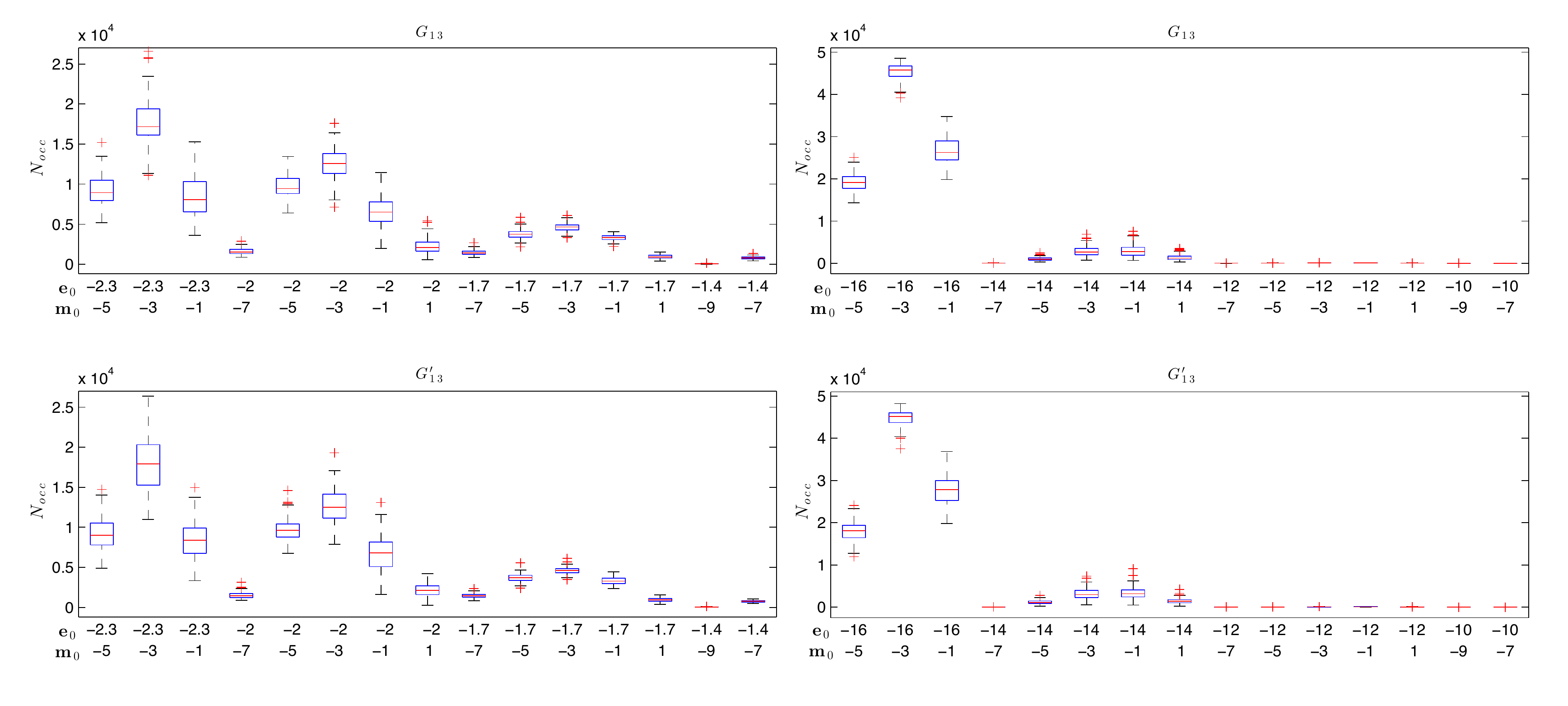} 
\caption{The statistical distribution of measurement outcomes $N_{occ}$ on the pairs $\{G_{13}%
, G_{13}^{\prime}\}$ obtained by averaging over $100$ cycles for each of the $100$ different embeddings considered ($10000$ programming cycles in total). The horizontal red line
corresponds to the median of the data while the edges of the blue boxes correspond to the
$1$st and $3$rd quartile. Each red cross is an outlier measurement. The outcomes have been filtered after choosing the pair of classical observables \{$\mathbf{e}_{0}$, $\mathbf{m}_{0}$\}. Data showed in the left  panels correspond to the choice $J=h=1/7$. In the right panels  $J=h=1$, that  is the maximum strength of the couplings allowed by the hardware. With the given choice of classical observables, the distribution of measurement outcomes is not able to distinguish the two graphs, nor at the classical, neither at the quantum level.}%
\label{data13a}%
\end{figure}
\begin{figure*}
\centering
\includegraphics[width=0.95\columnwidth]{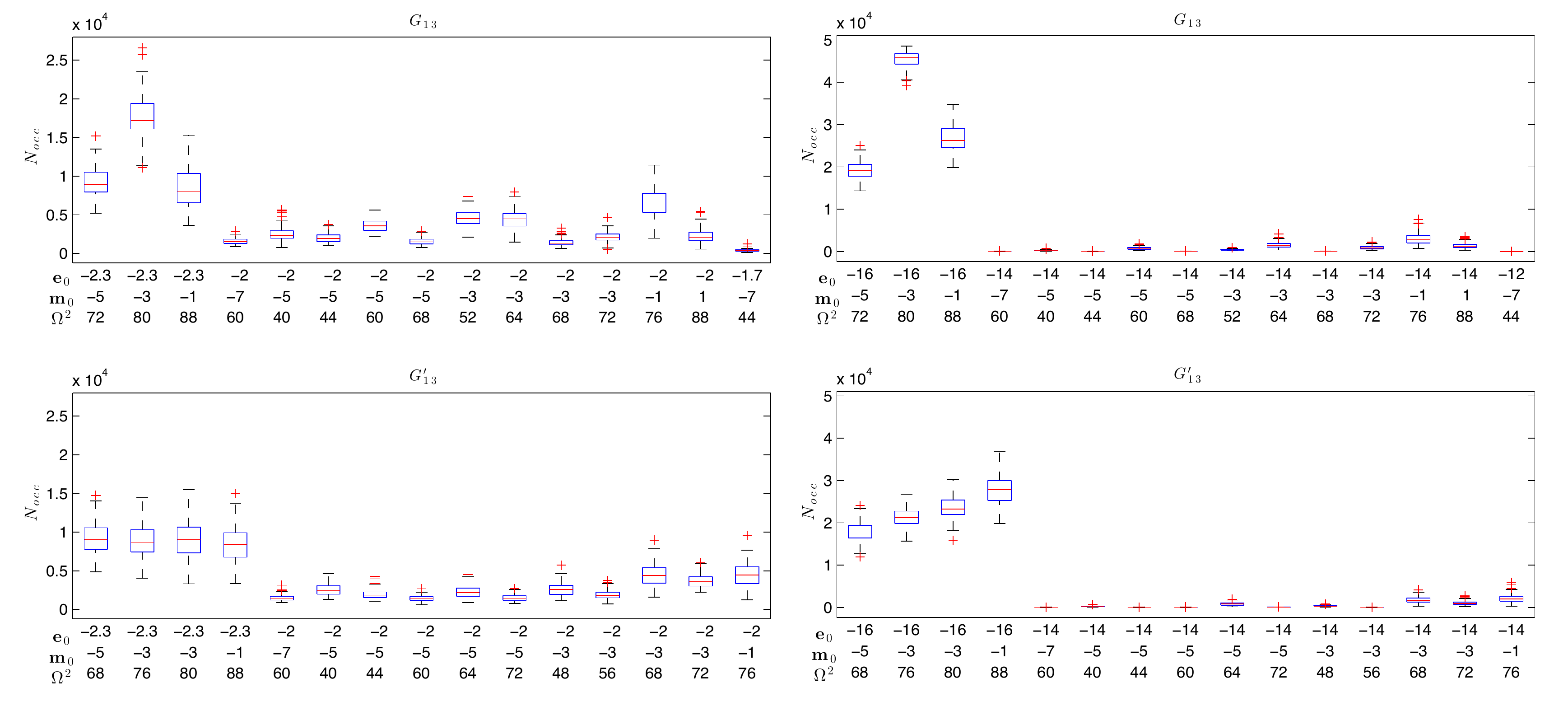} 
\caption{The statistical distribution of measurement outcomes $N_{occ}$ on the pairs $\{G_{13}%
, G_{13}^{\prime}\}$.  The outcomes have been now filtered after choosing a triplet of classical observables  
 \{$\mathbf{e}_{0}$, $\mathbf{m}_{0}$, $\Omega^{2}$\}. Data showed in the left  panels correspond to the choice $J=h=1/7$. In the right panels  $J=h=1$. Using a third observable distinguishes the two graphs at the classical level.}%
\label{data13b}%
\end{figure*}
\begin{figure*}
\centering
 \includegraphics[width=0.95\columnwidth]{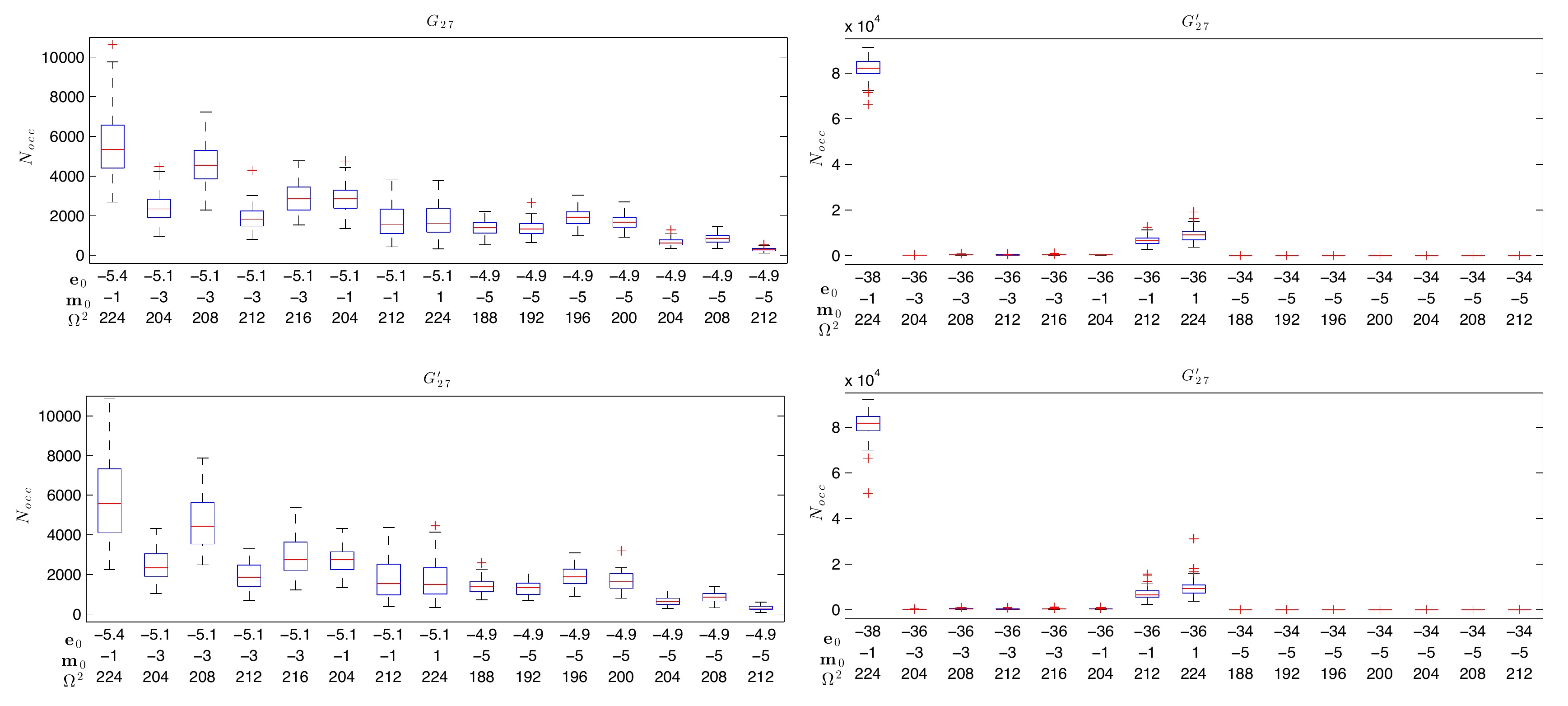} 
\caption{The statistical distribution of measurement outcomes $N_{occ}$ for the pairs
$\{G_{27},G_{27}^{\prime}\}$. $J=h=1/7$ in the left panels. $J=h=1$ in the right panels. Filtering the outcomes after choosing the triplet   
 \{$\mathbf{e}_{0}$, $\mathbf{m}_{0}$, $\Omega^{2}$\}  does not distinguishes the graphs at the classical level, and the introduction of additional observables is needed. The shape of the two distributions is also the same, meaning that the two graphs are not distinguished at the quantum level either.}%
\label{data27}%
\end{figure*}
{\bf Experiments.} Disregarding computational complexity aspects,
we have highlighted that from the theoretical point of view one can
hear the shape of certain quantum Ising models, while it is not
possible for the classical analogue. We subsequently encode on the
same physical system pairs of non-isomorphic graphs that are
co-spectral, longitudinal field co-Ising (and consequently co-Ising),
but not quantum co-Ising. Rather remarkably, our set up finds an experimental 
implementation in the optimization technique called \emph{quantum
  annealing}~\cite{Finnila,Kadowaki,Brooke99,Santoro92}. In this
technique, the system evolves adiabatically according to the following time-dependent Hamiltonian%
\begin{equation}
H_{QA}(G,J,h,\Delta,s)=sH_{L}(G,J,h)+(1-s)\Delta M_{T}, \label{qa}%
\end{equation}
where $s=t/T_{tot}$; $t$ is a time parameter and $T_{tot}$ is the total
duration of the dynamics. At the beginning of the computation, the system is prepared in the  ground state of the initial
simple Hamiltonian $H_{QA}(G,J,h,\Delta,0)=\Delta M_{T}$. On the basis of the adiabatic theorem~\cite{Farhi00}, adiabatic quantum
annealing with general Hamiltonians has been shown to be a universal model of
computation by Aharonov \emph{et al.}~\cite{Aharonov07}. In synthesis, the
core idea is to evolve the system slowly enough towards a final ground state,
which is the solution of a computational task. While the success of this
paradigm depends on the ability of avoiding level crossings with \emph{ad hoc}
annealing schedules, Brooke \emph{et al.}~\cite{Brooke99} experimentally
observed that tunneling can hasten convergence to the solution.

In the setting specified by Eq.~\eqref{qa}, we are interested in measuring the
observables $\mathbf{e}_{0}$, $\mathbf{m}_{0}$, and $\mathbf{o}_{0}^{k}$. In a
realistic situation, temperature and environment will usually excite the
system. While these effects are disruptive in the standard applications of
quantum annealing, we regard such a non-ideal implementation as a way to
generate the statistics of low energy states on which we measure the
corresponding observables. For this purpose, we run experiments on a D-Wave
Vesuvius\ programmable annealer. The hardware consists of $503$ usable logical
bits\ on an integrated circuit with superconducting flux qubits (see~\cite{Johnson10}
for details on the technology). Quantum effects on the chip are currently
under investigation and there is evidence of quantum annealing on random spin
glass problems~\cite{Boixo12a,Boixo12b,SSSVa,SSSVb,Vinci}. The Hamiltonians that can be realized with the
device are exactly of the type in Eq.~\eqref{qa}, where $s$ is a non-linear
function of time. The most general form of the final Hamiltonian $H_{L}$ is given by
an Ising model whose possible spin interactions are constrained by the chip
architecture. A particular limitation of the hardware is that measurements can
be performed only at the end of the evolution. Thus, the maximal information
that we can extract is encoded in the multivariate polynomials of Eq.~\eqref{mu}. On the other hand, the final state of the chip is a result of a
dynamics also governed by the transversal field $M_{T}$. In fact, our experiments attempt to identify the effects of $M_{T}$ in the final statistics after the measurement outcomes are filtered out using various type of multivariate polynomials.

We have tested the annealer on two pairs of non-isomorphic graphs $G$ and
$G^{\prime}$ ($G_{13}$ and $G_{13}^{\prime}$, $G_{27}$ and $G_{27'}^{\prime}$, in Supplementary Information Section B and C) such that $S_{A(G)}%
=S_{A(G^{\prime})}$, $S_{H_{L}(G,J,h)}=S_{H_{L}(G^{\prime},J,h)}$, and
$S_{H_{T}(G)}\neq S_{H_{T}(G^{\prime})}$, \emph{i.e.}, with equal spectra of the adjacency matrix, equal classical
spectra, even with a longitudinal field, and different quantum spectra.  To illustrate a possible (arbitrary)\ refinement as introduced
by Eq.~\eqref{mu}, we include an extra observable, $\Omega^{2}$, corresponding to
the next-nearest neighbor interaction energy:\ $\Omega^{k}=\sum
_{i,j}H(i,j)[A^{k}(G)]_{i,j}$. Notice that $H(G)=\Omega^{1}/2$. Figure~\ref{data13a} shows the statistics of measurement outcomes when the states are distinguished through the doublet of observables 
 \{energy, magnetization\}  on the pairs $\{G_{13}, G_{13}^{\prime}\}$, for $J=h=1/7$ and $J=h=1$. These are respectively the smallest and the largest values that can be reliably set on the hardware. The final states are organized according to the values of the two observables. As a consequence of the fact that the two graphs are co-Ising, the measured values of the pairs \{energy, magnetization\} are the same, and cannot be used to distinguish the two graphs. Moreover, the shape of the two distributions is also the same up to statistical errors. The shape of this distribution is assumed to be a
consequence of (noisy) open system quantum
dynamics~\cite{Boixo12a,Boixo12b} (see Supplementary Information Section D for a comparison between experimental and thermal statistics). This means that we are not able to identify differences in the final distributions that may arise due to the different quantum spectra, \emph{i.e.} due to non-equivalent quantum evolution along the annealing schedule.
 
  The graphs are indistinguishable by
measuring energy and magnetization only. However, they become
distinguishable in Figure~\ref{data13b} by measuring the triplet \{energy, magnetization,
$\Omega^{2}$\}, as clearly visible in the statistics obtained
with the chip.  The pair $\{G_{27}, G_{27}^{\prime}\}$ is not
distinguished by the triplet on the experimental data, as showed in Figure~\ref{data27}. It should be possible, in principle, to classically distinguish these graphs with the introduction of additional observables. Similarly to what happens for the $G_{13}$ pair, there are no noticeable differences in the shape of the final distributions that can detect differences in the quantum spectra.

\section{Conclusions}

The interplay between combinatorics and the
classical Ising model is well-established. We have introduced a
general family of physically meaningful graph polynomials suggesting a
hierarchy of graph invariants. We have demonstrated that the quantum
Ising model is a finer sieve to distinguish graphs than its classical
analogue by considering the quantum partition function as a graph
invariant. We have tested experimentally its distinguishability power
on a D-Wave programmable annealer, by taking graphs with different
quantum spectra and the same classical Ising partition function. We
used the hardware unconventionally to generate the statistics of low
energy eigenvalues rather than focusing on the ground state. The data
obtained can distinguish one pair of graphs when measuring with
respect to a classical refinement of the partition function. We did not find any measurable difference in the 
statistics of measurement outcomes of the two pairs that can be related to non-equivalent quantum dynamics.  Notice that
the transverse field spectra are very similar (Fig.~9 in the
Supplementary Material). Of course, differences expected in an ideal quantum
system are possibly lost due to decoherence when approaching the
classical regime at the end of the adiabatic evolution.

 Going beyond the scope of this work, it would be
interesting to compare the experimental data with numerical
simulations of the corresponding open quantum spin system at finite
temperature~\cite{Albash12}. We propose two approaches to amplify the
differences in the quantum spectra: (a) reduce substantially the
annealing time; (b) perform measurements when the transverse field is
on.  Both approaches require a modification of the current control of
the hardware.  Another interesting goal is to define other efficient
observables, such as $\Omega^2$, that would amplify the possible
differences in the measurement statistics. From the theoretical point
of view a natural open question is whether the transverse field alone
is sufficient to define a complete spectral graph invariant.

\section{Methods}

{\bf Experimental data collection.} In order to
collect enough statistics for averaging over biases and systematic errors, we
have considered $100$ embeddings in the chip for each graph. To average over
precision errors when setting the intended couplings on the machine, we have
run $100$ programming cycles for each embedding. For each cycle, we have performed $1000$
measurements. All the experiments have been performed choosing the shortest annealing time allowed by the hardware ($T_{tot} = 20\mu s$) in order to minimize the effects of thermal excitations.

\section{Acknowledgments}
We would like to thank Gabriel Aeppli, Andrew
Fisher, Andrew Green, Itay Hen, Daniel Lidar, Brent Segal, and Peter Young for
valuable discussion. This work has been done within a \textquotedblleft Global
Engagement for Global Impact\textquotedblright\ programme funded by
EPSRC, and with support from ARO grant number W911NF-12-1-0523, the
Lockheed Martin Corporation and DARPA grant number FA8750-13-2-0035.

\section{Author Contribution}
W.V., S.S. and P.A.W. provided the central ideas, that were further developed by all authors. K.M. performed the numerical exhaustive analysis of the graphs considered in the paper. W.V. performed all data collection and analysis. A.R. contributed in the data collection. S.B. and F.M.S. contributed in the data analysis. W.V.  and S.S. wrote the main manuscript text. All authors thoroughly reviewed the final manuscript.

\section{Additional Information}

{\bf Competing financial interests:} The authors declare no competing financial interests.


\newpage

\appendix

\section{Supplementary Material}

\section{{A}.{\emph{ Examples for Eq.~\eqref{exam} } }}
\begin{figure}[htbp]%
\centering
\includegraphics
{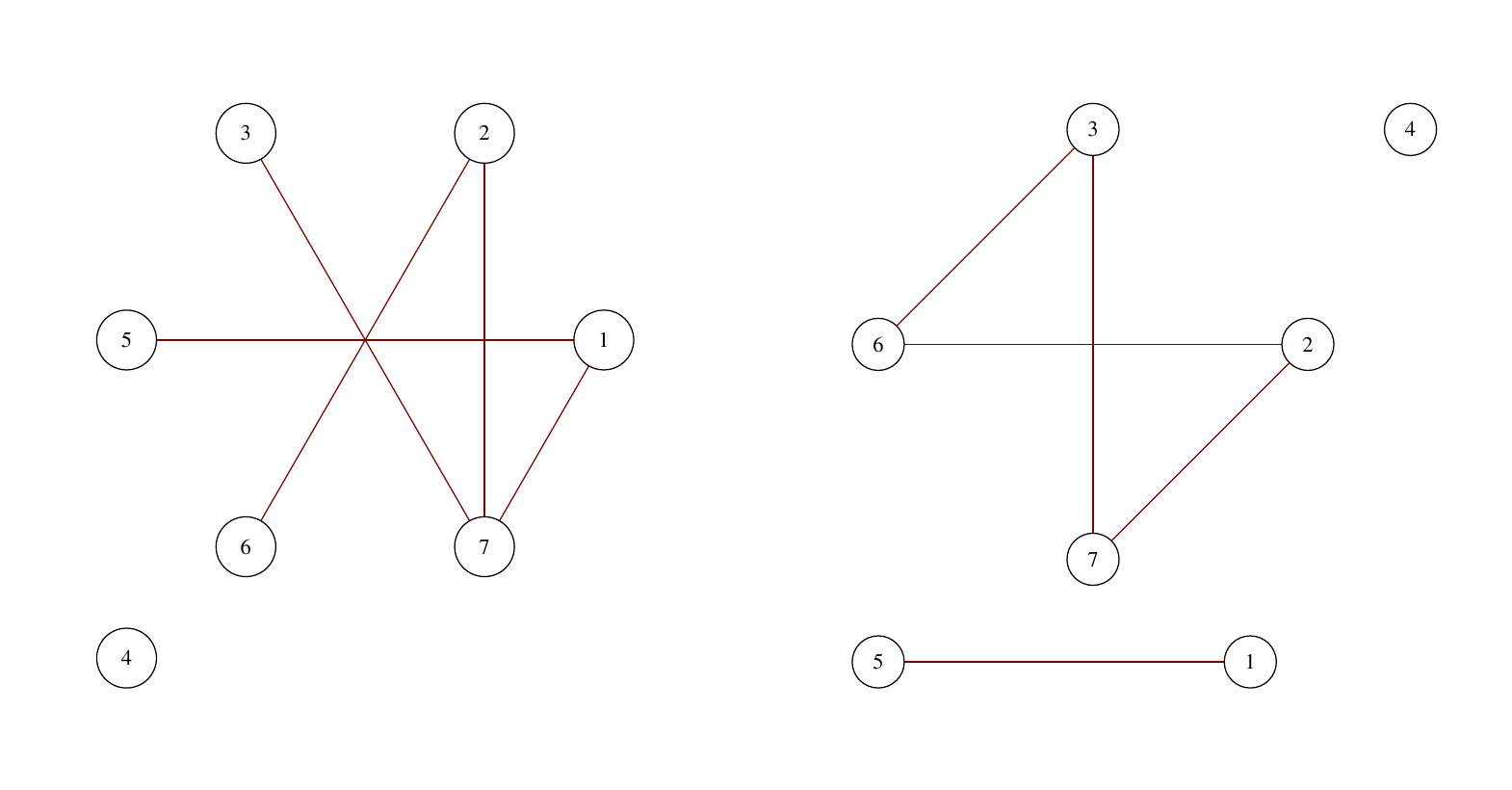}%
\caption{ Graphs $G_{1}$ (left) and $G_{2}$ (right) with $S_{H_{L}(G_{1})}=S_{H_{L}(G_{2})}$ and $S_{H(G_{1})}\neq
S_{H(G_{2})}$}%
\label{g1g2}%
\end{figure}

\begin{figure}[htbp]%
\centering
\includegraphics
{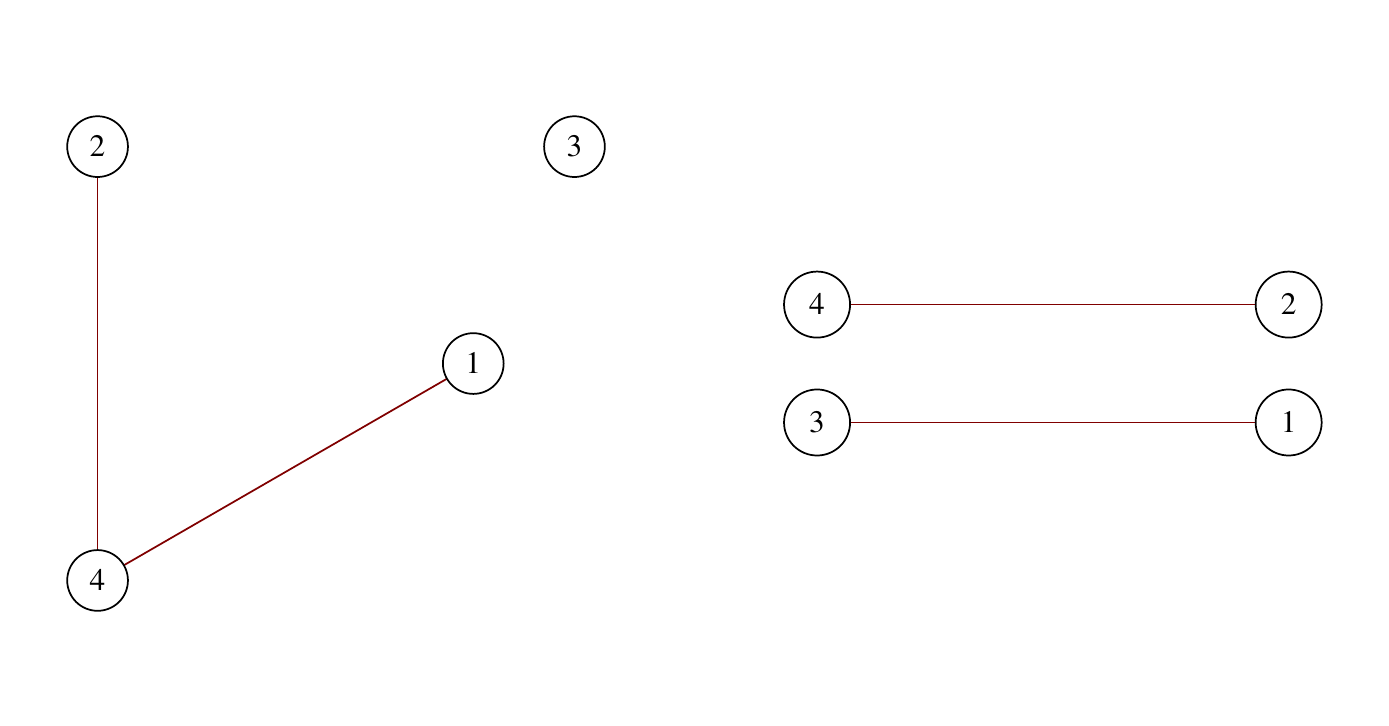}%
\caption{ Graphs $G_{3}$ (left) and $G_{4}$ (right) with
  $S_{H(G_{3})}=S_{H(G_{4})}$ and $S_{H_{L}(G_{3})}\neq S_{H_{L}
    (G_{4})}$.}%
\label{g3g4}%
\end{figure}

The two graphs
$G_{1}\ncong G_{2}$ on $7$ vertices with
\begin{align}
E(G_{1}) &  =\{1,5;1,7;2,6;2,7;3,7\},\nonumber\\
E(G_{2}) &  =\{1,5;2,6;2,7;3,6;3,7\},
\end{align}
are such that $S_{H_{L}(G_{1})}=S_{H_{L}(G_{2})}$ and $S_{H(G_{1})}\neq
S_{H(G_{2})}$ (see Fig.~\ref{g1g2}). We note that in general $S_{H_{L}(G_{1}%
,J,h)}\neq S_{H_{L}(G_{2}^{\prime},J,h)}$. The two graphs $G_{3}\ncong G_{4}$ on $4$ vertices with
\begin{align}
E(G_{3}) &  =\{1,4;2,4\},\nonumber \\
E(G_{4}) &  =\{1,3;2,4\},
\end{align}
are such that $S_{H(G_{3})}=S_{H(G_{4})}$ and $S_{H_{L}(G_{3})}\neq
S_{H_{L} (G_{4})}$ (see Fig.~\ref{g3g4}).

\bigskip

\section{{B}.{ }\emph{Experiments for }$G_{13}$\emph{ and }%
$G_{13}^{\prime}$ \emph{} }
\begin{figure}[htbp]%
\centering
\includegraphics
{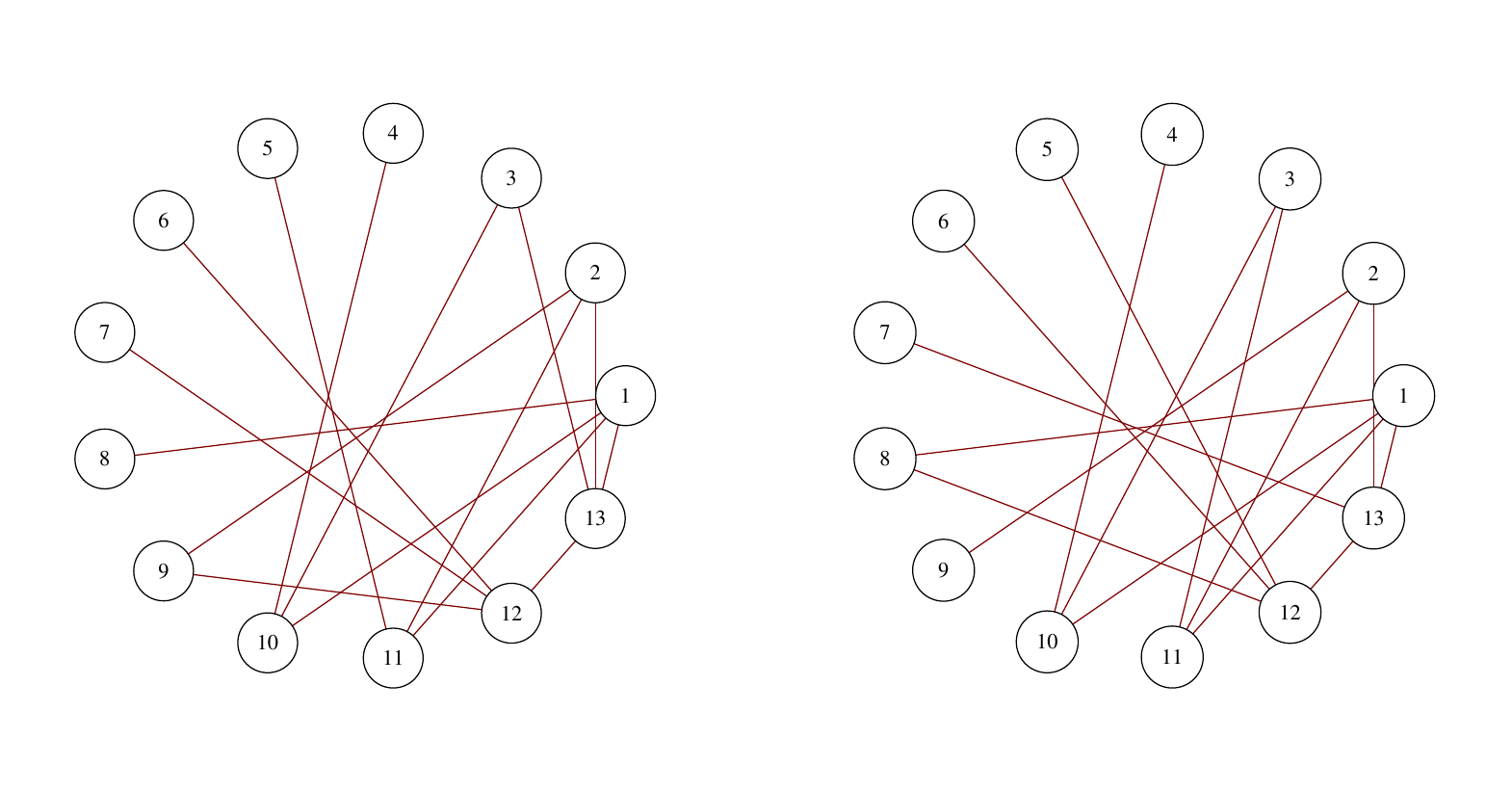}%
\caption{ Graphs $G_{13}$ (left) and $G_{13}^\prime$ (right).}%
\label{g13g13p}%
\end{figure}

The two graphs $G_{13}\ncong G_{13}^{\prime}$ on $13$
vertices with%
\begin{align}
E(G_{13})  &
=\{1,8;1,10;1,11;1,13;2,9;2,11;2,13;3,10;3,13;4,10;5,11;6,12;7,12;9,12;12,13\},\nonumber \\
E(G_{13}^{\prime})  &
=\{1,8;1,10;1,11;1,13;2,9;2,11;2,13;3,10;3,11;4,10;5,12;6,12;7,13;8,12;12,13\},\
\end{align}
are such that $S_{A(G_{13})}=S_{A(G_{13}^{\prime})}$, $S_{H_{L}(G_{13}%
,J,h)}=S_{H_{L}(G_{13}^{\prime},J,h)}$ and $S_{H_{T}(G_{13})}\neq
S_{H_{T}(G_{13}^{\prime})}$ (see Fig.~\ref{g13g13p}). The triplet \{energy, magnetization, $\Omega^{2}%
$\} distinguishes $G_{13}$ and $G_{13}^{\prime}$. The measurements performed
on the chip distinguish the graphs (see Fig.~\ref{data13a} in the main text).
Let $\lambda_{\max}(G)>\lambda_{2^{n}-1}(G)\geq\cdots\geq\lambda
_{2}(G)>\lambda_{\min}(G)$ be the eigenvalues of $H_{QA}(G,1,1,1,s)$ as a
function of $s$. Fig.~\ref{EIGS} shows $\lambda_{i}(G_{13})-\lambda_{\min
}(G_{13})$ and $\lambda_{i}(G_{13}^{\prime})-\lambda_{\min}(G_{13}^{\prime})$,
for $i=2,...,20$. The similarity between the two spectra is evident.

\begin{figure}[htbp]
\centering
\begin{tabular}{ccc}
\includegraphics[width=0.45\columnwidth]{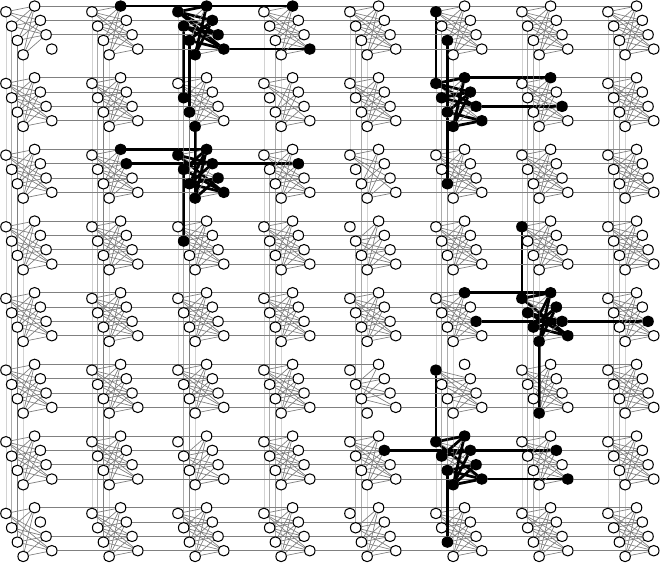}   & \qquad \qquad  &\includegraphics[width=0.45\columnwidth]{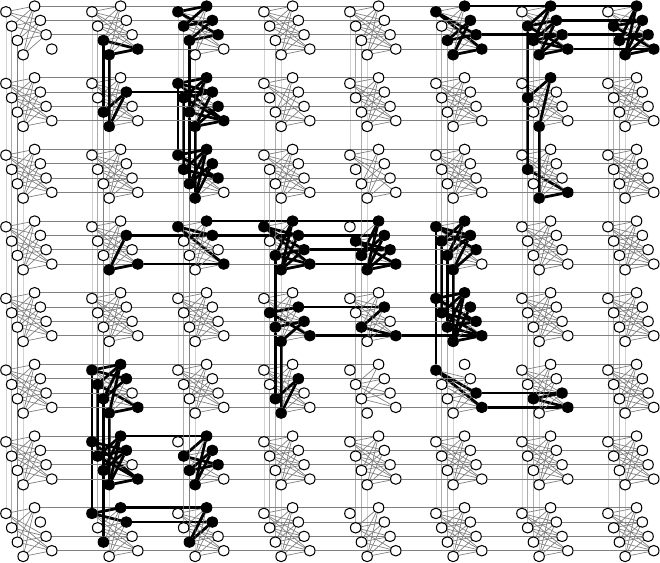}
\end{tabular}
\caption{The grid in the figure above shows the hardware connectivity. Each circle is a programmable superconducting pseudo spin, while the connecting lines represent the physical interactions that can be implemented in the chip. The black circles and lines correspond to the ``active" part of the hardware that is used to implement five example embeddings of $G_{13}$ (left)\ and $G_{27}$ (right) on the chip. }%
\label{Emb}%
\end{figure}

 Fig.~\ref{Emb}
contains five embeddings of the graph $G_{13}$ (and $G_{27}$) on the chip. We include this
figure to clarify the notion of physically different
embeddings of the same graph.

\bigskip

\section{\emph{C}.{ }\emph{Experiments for }$G_{27}$\emph{ and }%
$G_{27}^{\prime}$ \emph{}}
\begin{figure}[htbp]%
\centering
\includegraphics[width=\textwidth]
{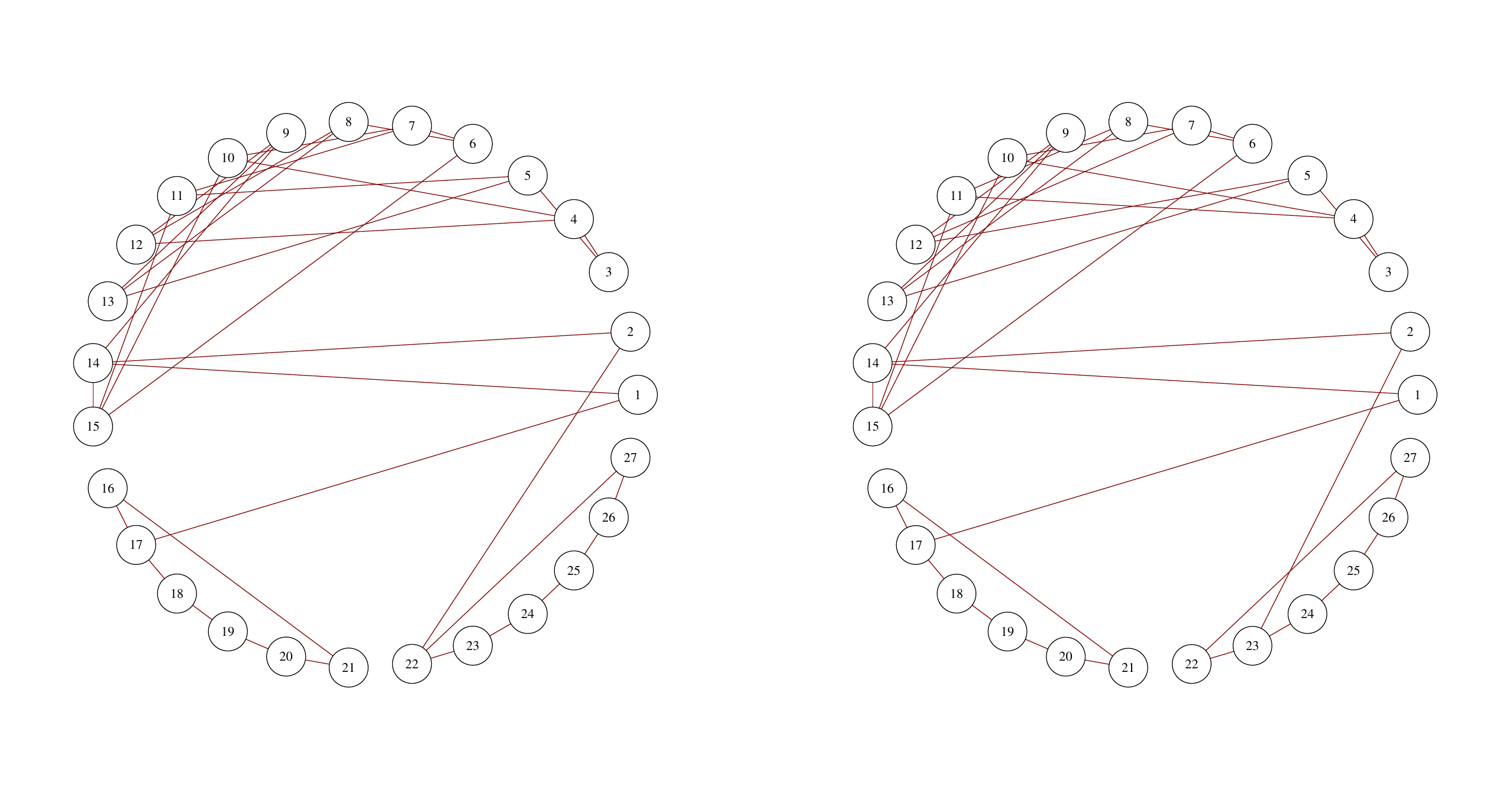}%
\caption{ Graphs $G_{27}$ (left) and $G_{27}^\prime$ (right).}%
\label{g27g27p}%
\end{figure}

\begin{figure}[htbp]%
\centering
\includegraphics[
width=5in
]%
{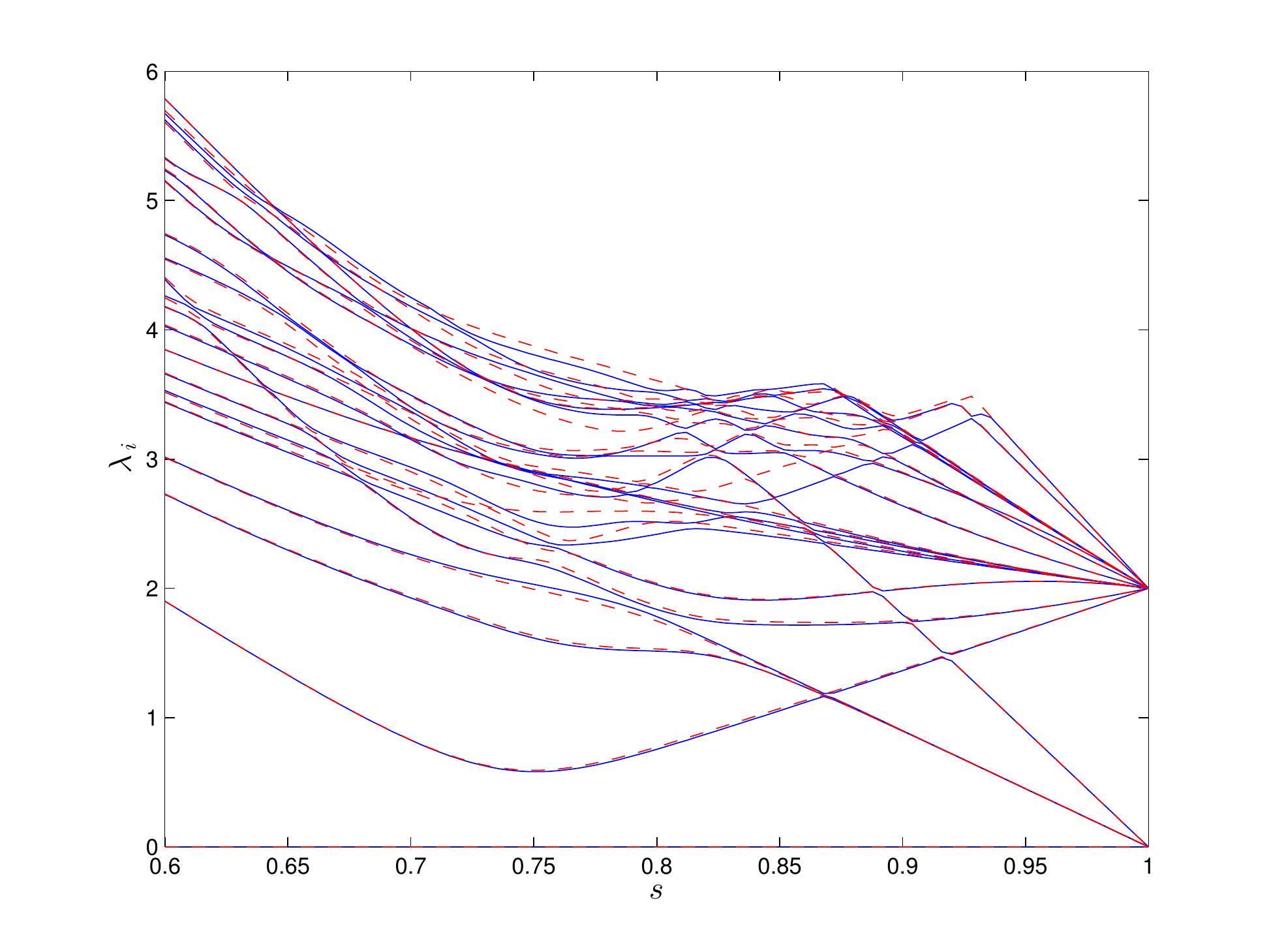}%
\caption{Quantum spectra $\lambda_i$ of $H_{QA}(G_{13},1,1,1,s)$ (solid line)\ and
$H_{QA}(G_{13}^{\prime},1,1,1,s)$ (dotted line)\ as a function of $s$. }%
\label{EIGS}%
\end{figure}

Consider the two graphs $G_{27}\ncong G_{27}^{\prime}$\ with
edges%
\begin{align}
E(G_{27})  &
=\{1,14;1,17;2,14;2,22;3,4;3,5;4,3;4,10;4,12;5,3;5,11;5,13;6,7;6,8;6,15;\nonumber \\
& 7,10;7,11;8,13;9,12;9,13;9,14;10,15;11,15;14,15;16,17;16,21;17,18;18,19;\nonumber \\
& 19,20;20,21;22,23;22,27;23,24;24,25;25,26;26,27\},\nonumber \\
E(G_{27}^{\prime})  &
=\{1,14;1,17;2,14;2,23;3,4;3,5;4,3;4,10;4,11;5,12;5,13;6,7;6,8;6,15;7,10;\nonumber \\
& 7,12;8,11;8,13;9,12;9,13;9,14;10,15;11,15;14,15;16,17;16,21;17,18;18,19;\nonumber \\
& 19,20;20,21;22,23;22,27;23,24;24,25;25,26;26,27\}.
\end{align}
For these graphs, $S_{A(G_{27})}=S_{A(G_{27}^{\prime})}$ and $S_{H_{L}%
(G_{27},J,h)}=S_{H_{L}(G_{27}^{\prime},J,h)}$ (see Fig.~\ref{g27g27p}). We do not know if
$S_{H_{T}(G_{27})}\neq S_{H_{T}(G_{27}^{\prime})}$. We would expect in this
case differences in the quantum spectra to be even smaller than those in the
case of $G_{13}$ and $G_{13}^{\prime}$. The triplet \{energy, magnetization,
$\Omega^{2}$\} distinguishes $G_{27}$ and $G_{27}^{\prime}$ when
considering the full spectrum. However, Fig.~\ref{data27} shows that the graphs are undistinguishable by the measurements
performed on the chip for $J=h=1/7$.  

\begin{figure}[htbp]
\centering
\begin{tabular}{cc}
\hspace{-0in}\includegraphics[width=0.5\columnwidth]{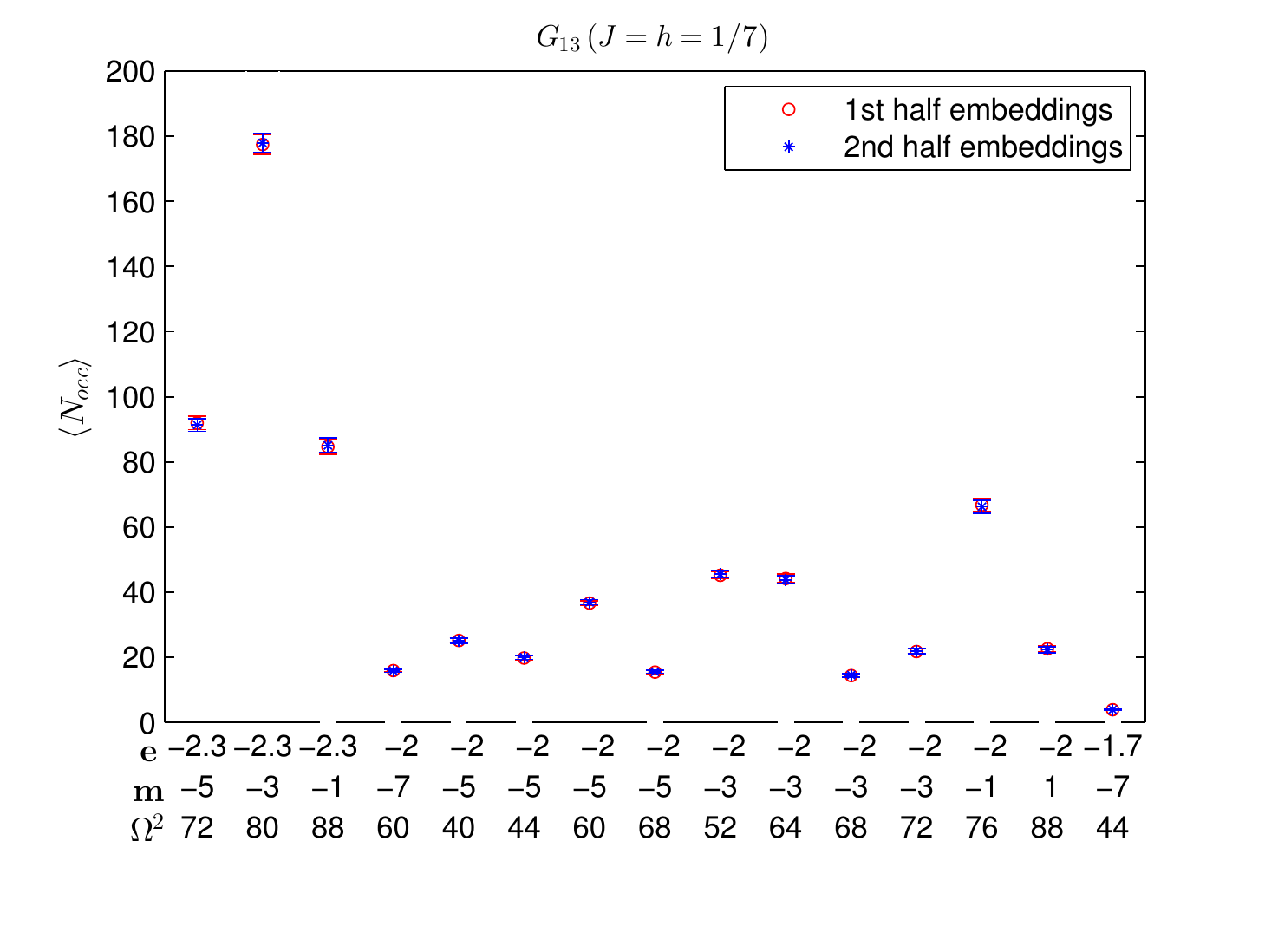}   &  \hspace{-0in}\includegraphics[width=0.5\columnwidth]{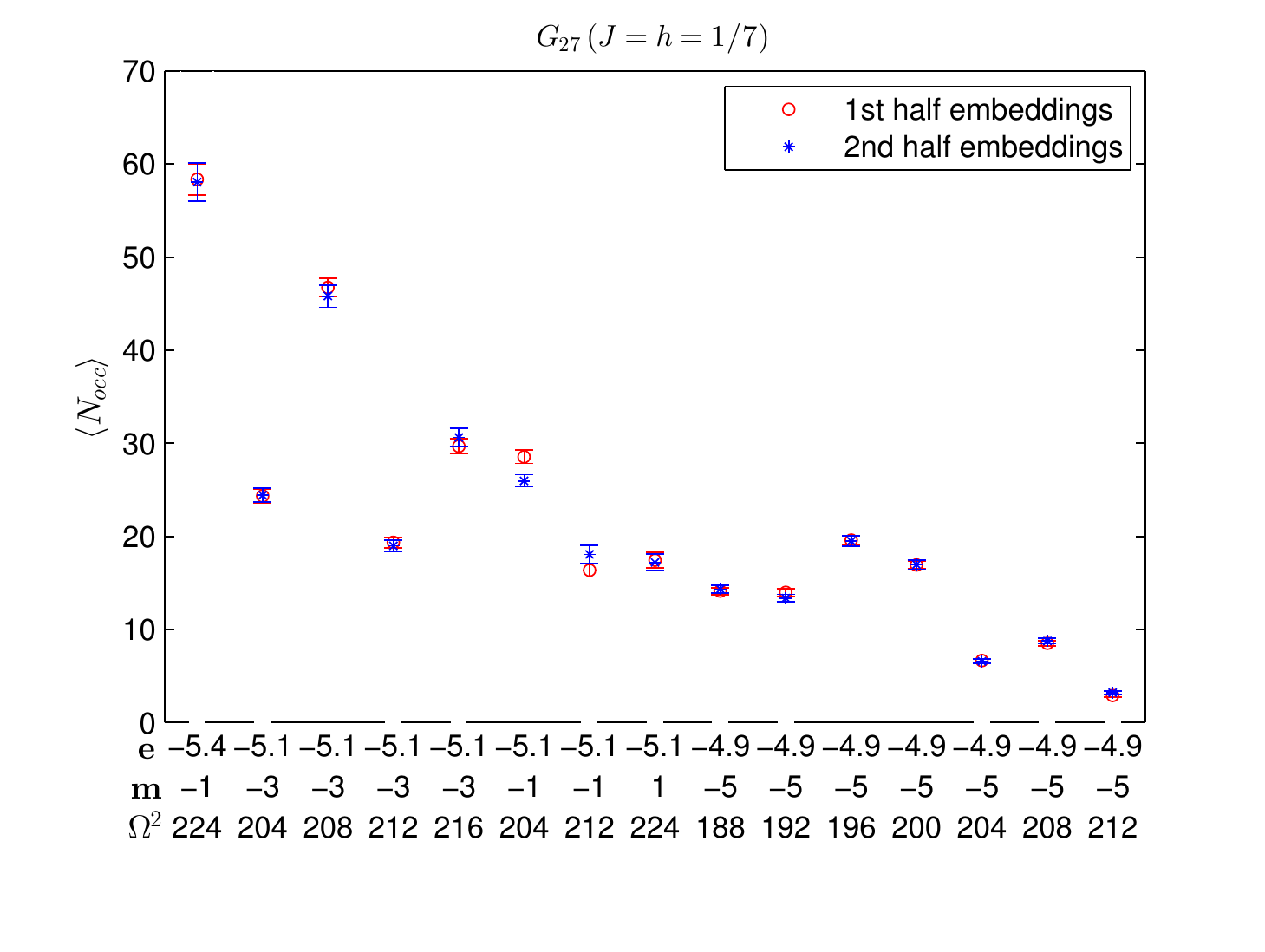}
\end{tabular}
\caption{Average value of measurement outcomes $\langle N_{occ} \rangle$ for each programming cycle for the graphs
$G_{13}$ (left) and $G_{27}$ (right). $J=h=1/7$. The statistics corresponding to one half of the embeddings
is shown in the red circles and the other half in the blue stars.  Error bars have been estimated with a bootstrap technique. We have extracted $5000$ bootstrap samples out of the two halves of the total programming cycles performed (10000). }%
\label{homogeneity}
\end{figure}

As a check that the data statistics are not critically dependent on
the embeddings considered, we show in Fig.~\ref{homogeneity} the
statistics corresponding to half of the embeddings in each case. 

\bigskip

\section{\emph{D}.{ }\emph{Comparison to Gibbs statistics} }
\begin{figure}[htbp]
\centering
\begin{tabular}{cc}
\hspace{-0in}\includegraphics[width=0.5\columnwidth]{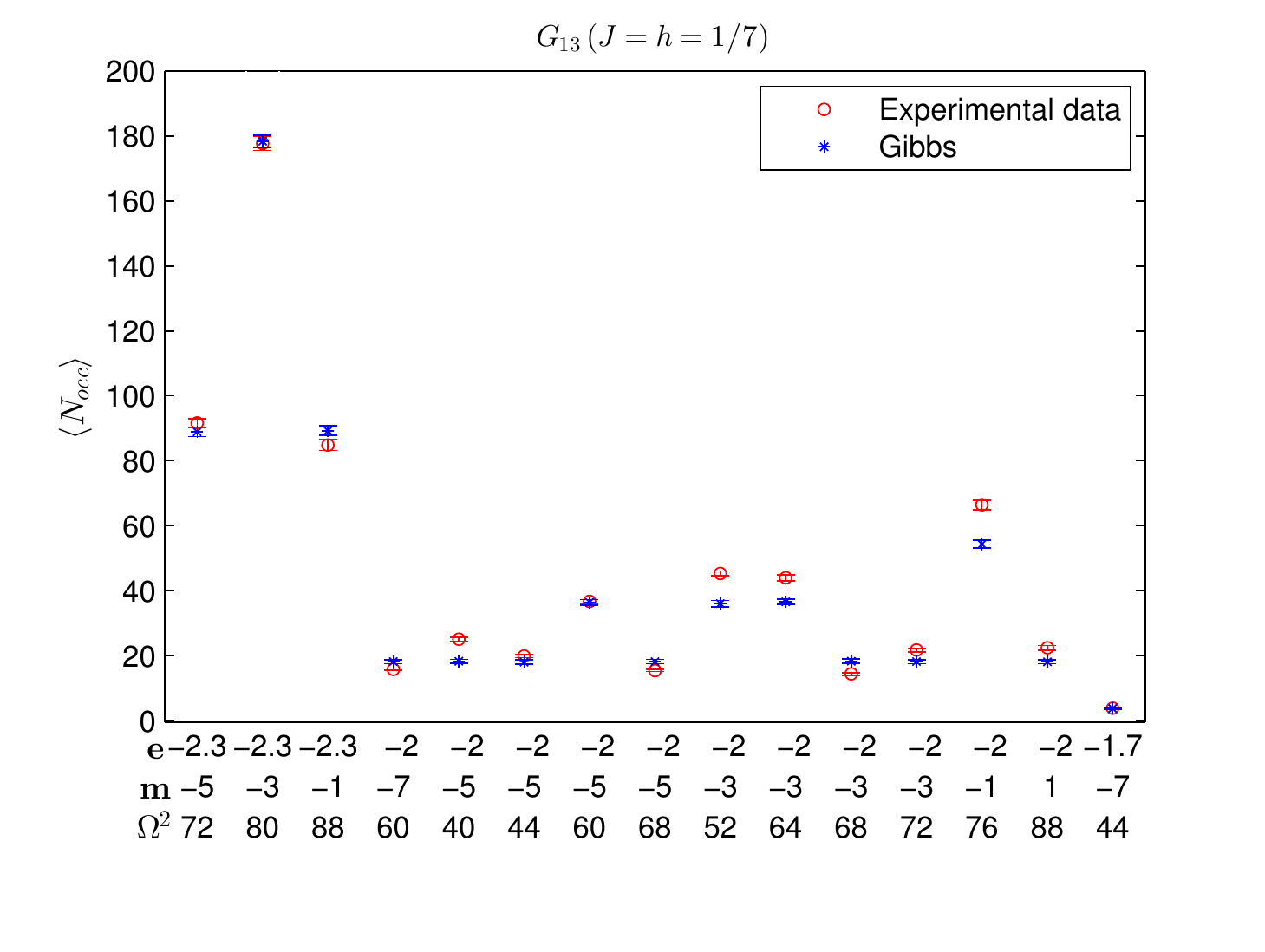}   &  \hspace{-0.in}\includegraphics[width=0.5\columnwidth]{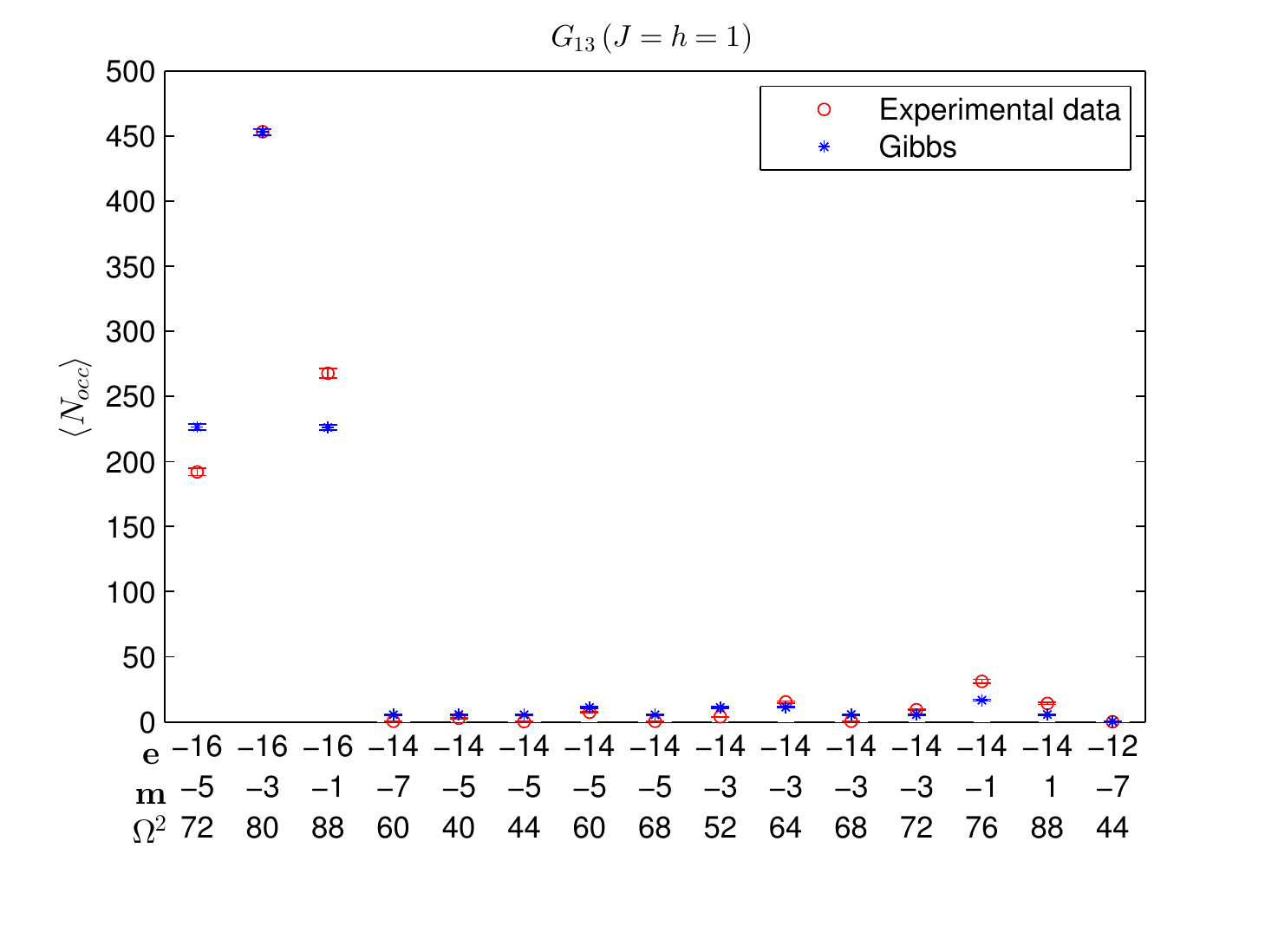}
\end{tabular}
\caption{The figure compares experimental data  to thermal Gibbs distributions for the graph $G_{13}$. The left panel corresponds to the upper left panel of Fig.~\ref{data13a} ($J=h=1/7$). The right panel corresponds to the upper right panel of Fig.~\ref{data13a} ($J=h=1$). The temperature to generate the Gibbs sampling has been chosen as the one that reproduces the statistics of the second collection of states.  Error bars have been estimated with a bootstrap technique. We have extracted $1000$ bootstrap samples out of $1000$ randomly picked programming cycles. The numerically generated Gibbs statistics contains the same number of generated states ($10^6$). We have checked that the error estimates are consistent by repeating the bootstrap sampling.}%
\label{Gibbs}
\end{figure}

To substantiate  our assumption that the  
experimental data are a consequence of a noisy quantum evolution, we have compared the signature distributions to those obtained by sampling from a thermal Gibbs distribution defined on the classical Ising models associated to the graphs. The results are shown in Fig.~\ref{Gibbs} for the $G_{13}$ graph and two choices of the couplings. The temperature for the Gibbs sampling has been chosen to best reproduce the experimental data.  The two distributions are notably different, although they show many qualitative common features.

%
%

\end{document}